\documentclass[12pt,preprint]{aastex}

\begin{document}
\title{
Super-solar Metallicity in the NLS1 Galaxy Markarian\,1044
}
\author{Dale L. Fields\altaffilmark{1}, Smita Mathur\altaffilmark{1}, Richard W. Pogge\altaffilmark{1}, Fabrizio Nicastro\altaffilmark{2}, Stefanie Komossa\altaffilmark{3} \& Yair Krongold\altaffilmark{2,4}}
\altaffiltext{1}
{Department of Astronomy, the Ohio State University, 140 West 18th
Avenue, Columbus, OH 43210, USA}
\altaffiltext{2}
{SAO,
60 Garden Street, 02138, Cambridge, MA, USA}
\altaffiltext{3}
{Max-Planck-Institut f\"ur extraterrestrische Physik,
Giessenbachstrasse 1, D-85748 Garching, Germany}
\altaffiltext{4}
{Instituto de Astronom\'ia Universidad Aut\'onomica de M\'exico, Apartado Postal 70 - 264, Ciudad Universitaria, M\'exico, D.F., CP 04510, M\'exico}

\email{fields@astronomy.ohio-state.edu}

\begin{abstract}

The determination of the bulk metallicity and the abundance mixture of
various elements is very difficult in quasars and AGNs because only a
few lines are observed and the ionization correction is unknown. Most
abundance studies of AGNs assume the N/C ratio scales as metallicity
(nitrogen goes as metallicity squared) and so serves as a metallicity
indicator.  We present an accurate metallicity determination of the
narrow-line Seyfert 1 (NLS1) galaxy Markarian 1044, using \ion{O}{6}
column density measurements from the {\it Far Ultraviolet Spectroscopic
Explorer} (FUSE) together with \ion{C}{4}, \ion{N}{5}, and \ion{H}{1}
from {\it Hubble Space Telescope} observations.  In this absorption line
study we find that the circumnuclear gas in Mrk\,1044 has a metallicity
of at least five times solar. This is consistent with the expectation
that NLS1s have a high metallicity, similar to that found in
high-redshift quasars.  More surprisingly, we find that the absorbing
material requires a near-solar mixture.  In other words, the N/C is
consistent with the solar ratio, and does not scale with the
metallicity. This suggests that the chemical enrichment scenario for
this object, and perhaps for AGNs in general, may be different from the
traditional model of galactic metal enrichment, at least in the
high-metallicity regime.

\end{abstract}


\section{Introduction} \label{sec:intro}

The study of metallicity, or more precisely, the study of metallicity
indicators in AGNs has a long history (e.g. \citealt{BK69}).  The
metallicity of material in an AGN has the potential to tell us about the
star formation history without looking at the stars themselves, which is
very important in such objects as high-redshift quasars where what
appears to be solar or above metallicity is thought to be achieved in
what must be less than a Gyr.  Metallicity studies have also been
conducted of the local AGNs, the less powerful Seyferts.  While some may
consider the enrichment questions less fundamental in these type of AGNs,
there are still curious correlations between the metallicity of gas
(host) and the luminosity (AGN) \citep{SN}.  While standard Seyferts may
have perfectly reasonable metallicity indicators for their host, there
are some local AGNs with what appears to be unusually high metallicities:
the Narrow-Line Seyfert 1.

The class of AGNs known as Narrow-Line Seyfert 1s (NLS1) are defined by
their relatively narrow permitted lines ($<2000$\,km\,$s^{-1}$) and a
weak [\ion{O}{3}]/H$\beta$ ratio ($<3$) \citep{OP}.  Their spectral
properties place them at one end of a distribution of AGN properties
known as Eigenvector 1 \citep{BG}.  Later studies of their X-ray
properties found that their steep X-ray spectra compared to other
Seyfert 1s \citep{BBF,BMF} continue to place them at one end of this
variation in AGN properties.

While it is not yet resolved, it is generally accepted that the physical
driver of Eigenvector 1 is the fractional Eddington accretion rate ($\rm
\dot{m}=\dot{M}/\dot{M}_{Edd}$) as suggested by \citet{P95} and
\citet{B02}.  This would set NLS1s as a class of AGNs that are undergoing
heavy relative accretion, in some ways similar to high redshift quasars.
Because NLS1s have a similar luminosity distribution as normal Seyfert
1s, this places the central black hole mass of NLS1s as smaller
(10$^6$-10$^7$) than those of standard Seyfert 1s ((10$^8$-10$^9$)
\citep[and references therein]{G04}.  Because there are relationships
between the mass/velocity dispersion of a spheroid and the mass of the
black hole it is host to \citep{G00, FM00, MF01}, this indicates that
NLS1s lie in galaxies with weak bulges.  This may not be completely true
as there is some evidence that NLS1s do not tend to lie on the standard
$M_{BH}-\sigma$ relation \citep{M01,GM04,MG05}.  NLS1s and high-redshift quasars share steep
X-ray slopes as compared to other AGNs \citep{G05}.  Another
difference between narrow-line and standard (broad-line) Seyferts is
that the emission line metallicity indicator \ion{N}{5}/\ion{C}{4} is
much stronger in NLS1s \citep{SN}.  If the \ion{N}{5}/\ion{C}{4} is to
truly a metallicity indicator, NLS1s should be placed along with high-luminosity
(high-redshift) quasars as the most metal rich of all AGNs.  While it is
not necessarily expected that NLS1s and QSOs have similar star formation
histories, there is no {\it a priori} reason why NLS1s and broad-line
Seyferts should share one.  It has been proposed that NLS1s are an
evolutionary phase by \citet{M2K} and if so, studying their metallicity
may inform us to the evolutionary properties of their host.  Even if
this hypothesis is invalid, and NLS1s are simply a distinct class of
medium-luminosity AGNs, then studying the metallicity is important to try
and distinguish why certain bulges host NLS1s instead of broad-line
Seyferts.

A problem lies, however, in whether the metallicity indicators studied
in fact return the correct metallicity.  Unlike in stars, where the physical conditions, input spectrum and localization of various lines are relatively understood, very little is truly well known about AGNs.  The paucity of lines and the fact that many of them may arise from unique regions of the AGN with what can be very different conditions does not assist our situation either.  Additionally, historically metallicity
studies have primarily used emission lines, for the simple reason that
these were all that were available or feasible.  And on an order of
magnitude basis, a stronger emission line should correspond to more of a
particular species.  Line ratios should give information about the
relative quantities of particular elements.  Unfortunately this can be
muddied by the shape of the ionizing spectrum or lines coming from
physically separate regions (just to name two problems).  Absorption
line studies, being insensitive to effects such as geometry or density, offer a better measure the quantity of a species.  While
absorption line studies are not feasible on survey scales, a few
detailed analyses of individual systems have the potential to provide a
sort of ``anchor'' to the much more widespread emission line
measurements.

In this paper we aim to define the relative (and absolute) abundances
for a few common elements in the NLS1 galaxy Mrk\,1044.  To this end,
this nearby AGN ($z=0.01645$) has been observed with the FUV
spectrograph onboard the {\it Far Ultraviolet Spectroscopic Explorer}
(FUSE) satellite.  Even with absorption line studies, however, the ionization correction remains a major problem in transforming the observed ionic column densities to a true abundance.  This problem is relatively common in any study that only
covers a small wavelength region and/or contains information on few
species of a particular element.  For this reason, we extend the baseline of observed species in Mrk\,1044 by adding \ion{O}{6} in the FUV range to measurements made in a STIS study at approximately
the same epoch \citep[henceforth F05]{hst}.  Neither dataset alone
could break the inherent degeneracies in making the photoionization
correction, but both datasets together select a small region of
parameter space from which we can determine the metallicity of this
system.  In \S\ref{sec:data} we detail the observations made and the
data reduction path followed.  In \S\ref{sec:analysis} we give the
methods with which we analyse the data.  We finish in
\S\ref{sec:conclude} with our conclusions.

\section{Observations and Data Reduction} \label{sec:data}

Mrk\,1044 was observed by the Far Ultraviolet Instrument onboard the Far
Ultraviolet Spectroscopic Explorer on UTC 2004 January 01 and 02.  The
observation was conducted over one $\approx7$ hour observation from
MJD 53005.74609375 to 53006.046875.  The data sets are D0410101.  The
received data was reduced by the standard CalFUSE pipeline.  The FUSE IDL tool FUSE\_REGISTER was used to
cross-correlate, weigh, and coadd the spectra.  We use the observations
with the 1A detector and the LiF mirror in this analysis.

\section{Analysis} \label{sec:analysis}
 
\begin{figure}
\epsscale{0.9}
\plotone{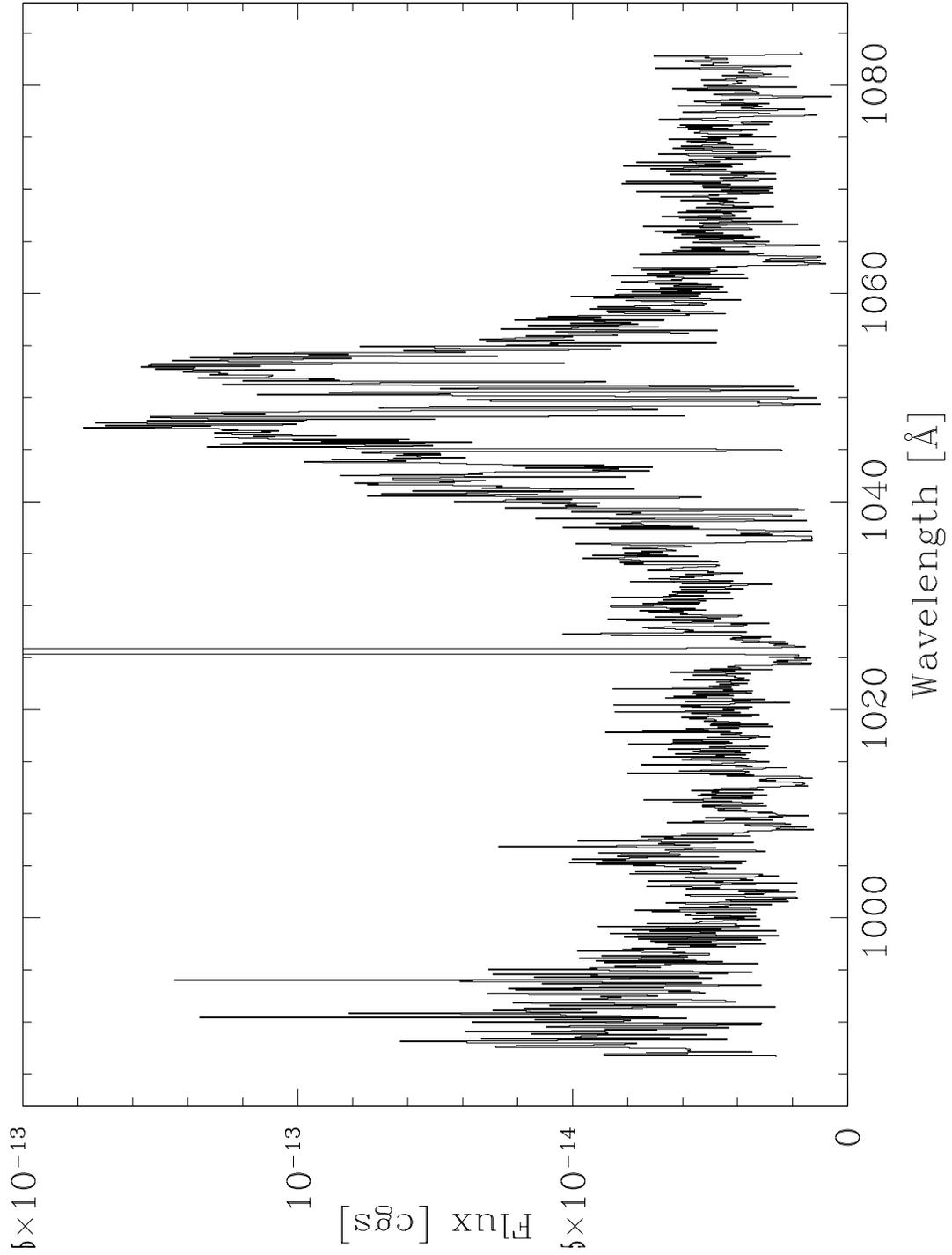}
\caption{\label{fig:fullspec}
Observed FUSE spectrum of Mrk\,1044.  Most prominent are
\ion{O}{6}$\,\lambda\lambda1032$,$1038$ and geocoronal Ly$\beta$
emission lines.  Less prominent are
\ion{N}{3}$\,\lambda\lambda990$,$992$ emission and a broad feature of
excess flux at the probable redshift of Ly$\beta$.  }\end{figure}

The UV spectrum observed by FUSE appears as most AGN spectra do, as can
be seen in Figure~\ref{fig:fullspec}.  The continuum is essentially
flat, with a slope indistinguishable from zero (expressed as
$\alpha=0.0$ in $f_{\nu}\propto\nu^{-\alpha}$).  The AGN imprints two
emission line complexes, the most obvious being the blend of Ly$\beta$
and \ion{O}{6}\,$\lambda\lambda1032$,$1038$ observed from 1030 to
1060\AA.  The other, much weaker, is
\ion{N}{3}\,$\lambda\lambda990$,$992$ observed around 1010\AA.  There
are also many absorption lines, most of which are Galactic in origin,
but some of which are the result of systems intrinsic to Mrk\.1044.

\begin{figure}
\epsscale{1.0}
\plotone{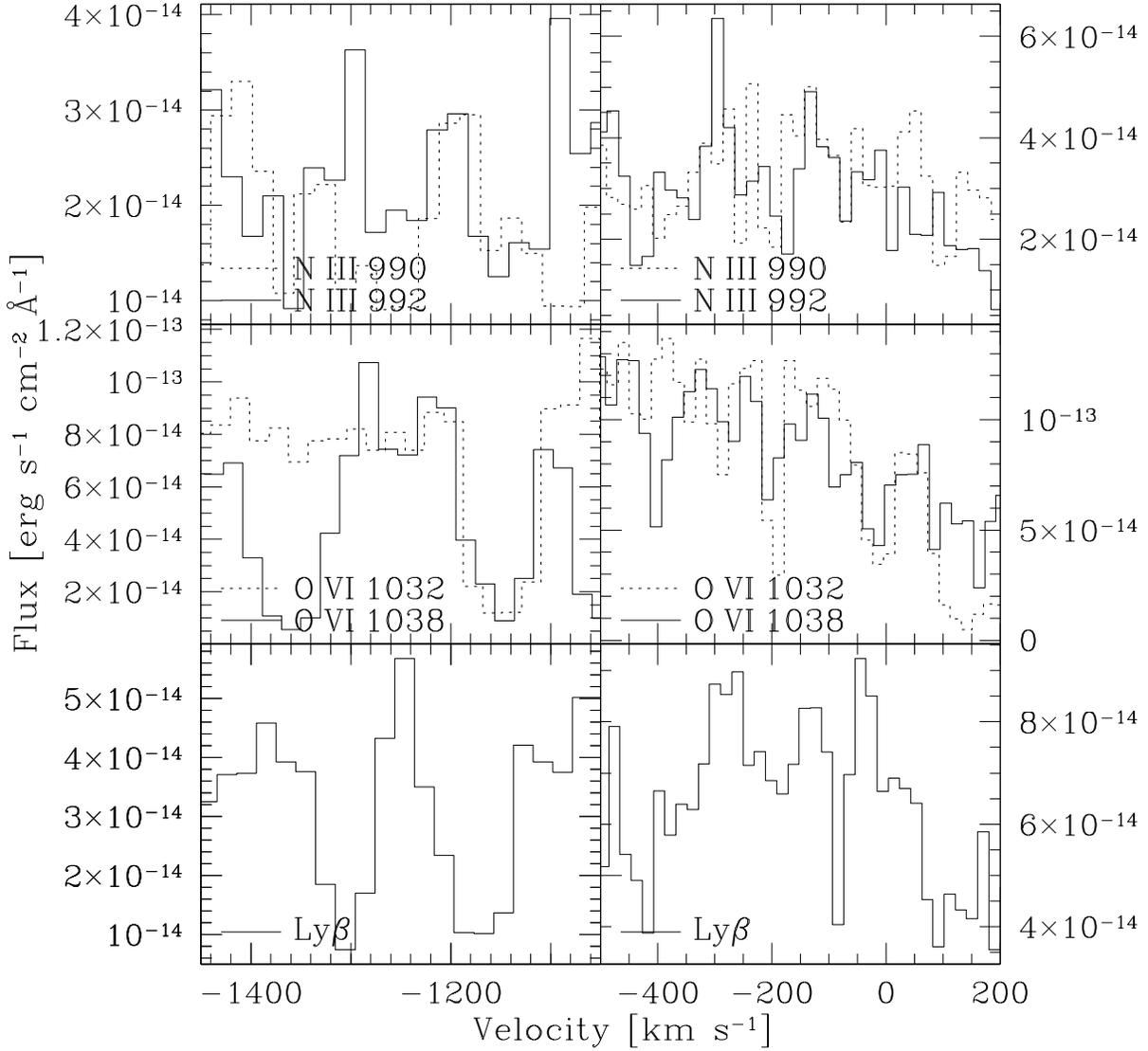}
\caption{\label{fig:velmap}
Velocity maps centered on the velocity of the absorption systems found
in F05.  System 1 (left) at $-1158$ and System 2 (right) at
$-286$\,km\,s$^{-1}$ match well with the velocities of $-1158$ and
$-286$\,km\,s$^{-1}$ found in F05.  The line seen in Lyman$\beta$ for
System 1 is a blend of Lyman$\beta$ and a Galactic H$_2$
line.}\end{figure}

The purpose of this paper lies in the absorption lines intrinsic to
Mrk\,1044, but this is complicated by the presence of so many Galactic
absorption lines.  As many absorption lines in an AGN spectrum are the
result of outflowing material with high relative velocities, line
identification must begin with locating known Galactic lines so as to
avoid contamination.  The Galactic source causing the most absorbing
features is molecular hydrogen (H$_2$) in the ISM.  We use the template used in
\citet{ark564} to model the H$_2$ absorption lines.  We then use line lists such as those in
\citet{galline} to identify the Galactic metal lines
\ion{C}{2}\,$\lambda\lambda1036$,$1037$, \ion{O}{1}\,$\lambda1039$,
\ion{Ar}{1}\,$\lambda1048$, and \ion{Fe}{2}\,$\lambda\lambda1055$,$1063$.
Then, guided by the results of F05, we constructed velocity maps, such
as can be seen in Figure~\ref{fig:velmap}.  We adopt the NASA/IPAC
Extragalactic Database (NED) value of the systemic velocity of Mrk\,1044
of 4932\,km\,s$^{-1}$.  This figure covers the same ranges as Figure 3
in F05, and shows what are likely the same absorbing systems as in Ly$\alpha$, \ion{N}{5}, and \ion{C}{4}, the strong System
1 at $-1158$ and the weaker System 2 at $-286$\,km\,s$^{-1}$.
Figure~\ref{fig:velmap} illustrates the danger of not fully identifying
Galactic lines before doing a velocity study.  The feature at
$\approx-200$ appears to be an absorbing system in all but
\ion{N}{3}\,$\lambda990$.  However, what appears to be the \ion{O}{6}\,$\lambda1032$ line
is really \ion{Ar}{1}\,$\lambda1048$.  In addition, the Ly$\beta$ at the
redshift of System 1 is heavily blended with a H$_2$ line at 1048\AA.
This figure also shows what is missing: no \ion{N}{3} absorption lines
are present at the positions of System 1 or 2, and no Ly$\beta$ is found
at System 2.  We note that there appears to be an absorption system at $v\approx0$\,km\,s$^{-1}$ relative to the systemic velocity of Mrk\,1044 in \ion{O}{6}.  The Ly$\beta$ spectrum at that velocity neither supports nor opposes such a determination.  However, F05's Figure 3 clearly shows that neither \ion{C}{4} nor \ion{N}{5} have absorption at that velocity, though Ly$\alpha$ is consistent with the presence of an absorber.  It is possible that this is a highly ionized system, but because we cannot extract reliable information about it due to its scarcity of absorption lines, we do not consider it further.

Measurements of the properties of these lines proved too difficult to
accomplish by modeling the emission and absorption lines along with a
H$_2$ template simultaneously in the STSDAS package SPECFIT
\citep{SPECFIT}.  This was mostly due to two factors: 1) the
unsuitability of gaussians to properly model the emission line profiles,
and 2) the inability of the H$_2$ template to successfully model the
Galactic absorption line complex.  Five gaussians were found to
adequately (but not satisfactorily) fit the blend of Lyman$\beta$ and
\ion{O}{6}.  However, the best fit solution was often not physically
acceptable, one example of which was the tendency of the two broad
\ion{O}{6} lines, when left unconstrained, to centroid themselves
asymmetrically (i.e. both ``inside'' or ``outside'' the two narrow
\ion{O}{6} lines).  The H$_2$ template failed to simultaneously fit both
the strong and weak lines.

\begin{figure}
\epsscale{0.8}
\plotone{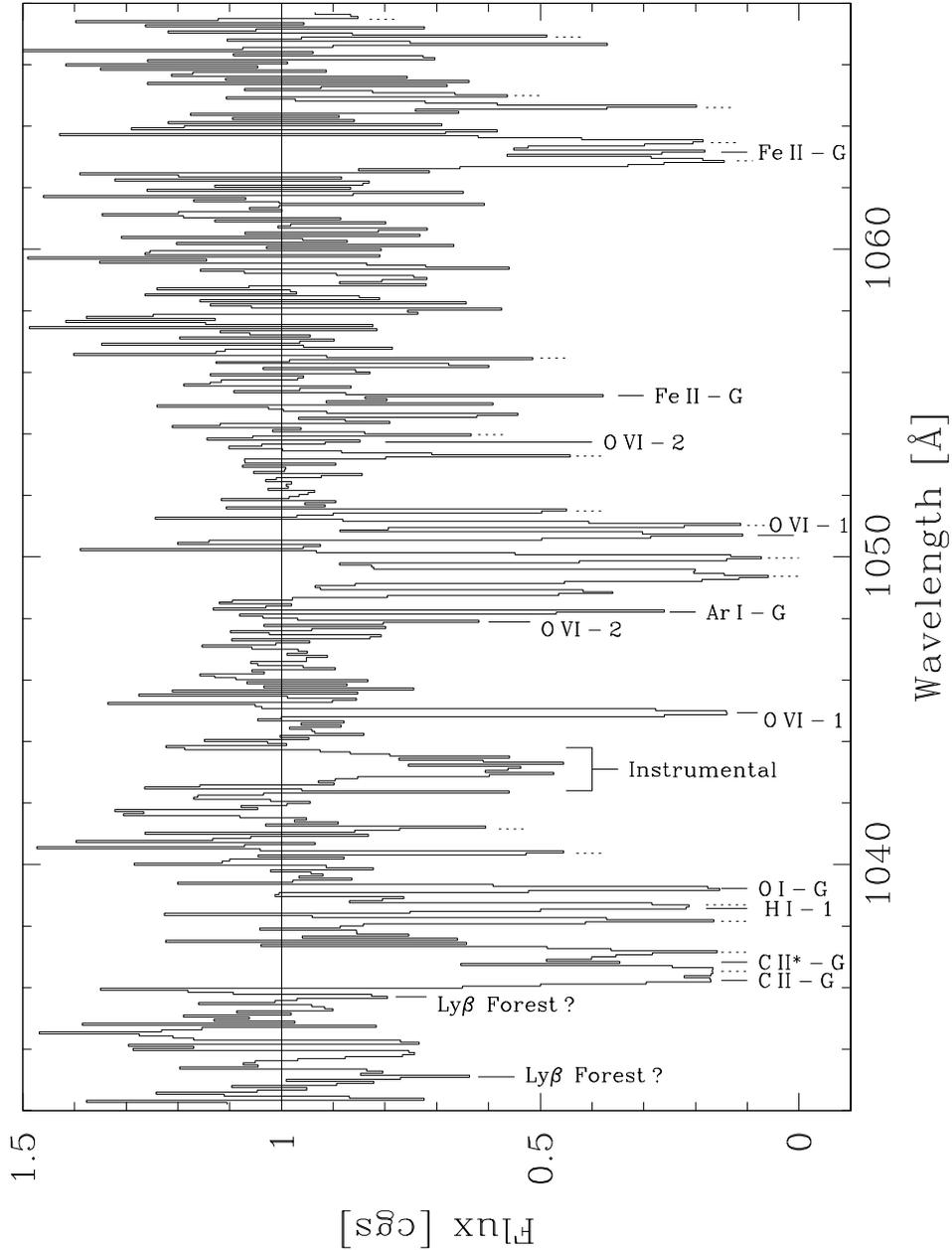}
\caption{\label{fig:flat}
One of the flattened spectra around the \ion{O}{6} emission line of
Mrk\,1044 used for our nine pseudo-continuum fits.  The H$_2$
subtraction has not been done with this spectrum.  Identified lines and
features are marked below the spectrum.  Lines intrinsic to Mrk\,1044
are marked by their system (1 or 2), lines belonging to Galactic sources
are marked with a 'G' and H$_2$ lines are marked with a dotted line.
The feature around 1043\,\AA\ is instrumental.  The \ion{C}{2} lines,
the Ly$\beta$ line for System 1, and the red \ion{Fe}{2} line are all
blended to some degree with H$_2$.}\end{figure}

Because the good measurement of these features with problems (emission
lines and H$_2$) are ancillary to the goal of well measuring the
intrinsic absorption systems, we attempt to remove the presence of these
problems one at a time.  To solve the issue of the emission lines, we
fit, by eye, a spline to the emission line structure and normalize the
spectrum by dividing by this ``pseudo-continuum.''  Such a normalized
spectrum can be seen in Figure~\ref{fig:flat}.  This spectrum is
obviously not perfectly fit everywhere (notice the poorly normalized
flux near 1059\AA), but does well enough around the positions of the
absorption lines we are measuring.  We then subtract an H$_2$ template
fit.  At this point, the lines intrinsic to Mrk\,1044 as well as the
Galactic metal lines can be fit.  Unfortunately, while normalizing the
spectrum worked well to solve the problem of the poorly fitting emission
lines, the issue of the H$_2$ fitting was never solved to our complete
satisfaction.  When the weaker H$_2$ lines were fit to, the template
predicts completely black cores for the strong lines which is not
reproduced in the data.  Also somewhat related to this is that the model's strong lines have too weak wings.  Increasing the model column density to
fit the wings of the data again produces black cores, and fitting to the
core of the strong lines ``creates'' two absorption lines at the wings
out of the residuals.  The blackness issue may be caused by the data.
As Figure~\ref{fig:flat} shows, all the strong absorption lines (most of
them due to H$_2$) reach a minimum around 0.15, which may indicate a
miscalibration in the flux.  However, the bottoms of all these lines are
already (individually) consistent with zero.  Also, resetting the flux
lower does nothing to solve the wings issue because all of these lines
lack multi-pixel troughs (which fitting to the wings predict).  Because
of this unresolved problem, we warn against trusting the H$_2$
subtraction, which will affect the fidelity of any lines molecular
hydrogen is blended with such as the Lyman$\beta$ line of System 1 as
well as Galactic \ion{C}{2} and \ion{C}{2}* at 1036 and 1037\,\AA.

We perform two separate analyses on the data.  The first is a standard
line fitting which gives us the observational parameters (equivalent
width, centroid, line width) and terminates in the estimation of the
column density via the curve of growth technique described in
\citet{Spitzer}.  The second directly integrates over the line structure
to give us the column density via the apparent optical depth technique
laid out in \citet{SS91}.  Comparing the results of these two separate
methods will give us additional insight into whether these lines give trustworthy results.  We feel confident in the utility of these methods even though studies such as \citet{A05} have found that some absorption features are inhomogeneous systems.  System 1 as shown in F05 is well sampled and is very well fit by a gaussian profile across the entire line.  The excellence of such a simple model fit implies that we are dealing with a kinematically simple system (a single absorber).  Unfortunately, System 2 is weak enough that it has much more noise in its profile, but it is not the focus of this investigation, nor does it figure into our conclusions in \S\ref{sec:conclude}.

The spectra are normalized in the analysis package LINER \citep{LINER}.
We also use LINER to derive an initial estimate of the properties of the
lines.  The resultant fits of this program were then used as the inputs
for the SPECFIT package.  This returns our final values of the
equivalent widths, line centroids, and line widths.  Equivalent widths
were then converted into column densities through the curve of growth
method.  We calculated the column density for velocity spread parameter
$b$-values of 10, 20, 40, and 80\,km\,s$^{-1}$, but we report only the
value derived from $b=20$ as that value gives consistent results between
the \ion{O}{6} doublet and was also the value used in F05.  The line centroiding is limited by localized detector distortions.  The galactic absorption lines show the absolute velocities should be shifted positively by about 3-5\,km\,s$^{-1}$, while the FUSE White Paper on the subject shows that the relative wavelength errors where our lines are located are about 8\,km\,s$^{-1}$.  We use this as our centroiding error.

To determine if the apparent optical depth method is appropriate, we
first test whether a line is resolved
($\Delta\lambda_{obs}\geq2\,\Delta\lambda_{lsf}$).  We compare FUSE's
instrument profile of $\approx20$\,km\,s$^{-1}$ with the line widths
returned from the SPECFIT package.  A stricter test is for the line to
be definitely resolved, in other words more than three standard
deviations above the nominal resolution limit.  If a line passes these
tests, we use the normalized residual flux ($I_r$) that results from the
LINER fit and integrate over the line.

We have determined two major sources of uncertainty in our line
parameters.  The first is the standard photon noise one expects and can
be propagated through each stage of the analysis process.  The second is
that associated with a sub-optimal fit to the pseudo-continuum.  We
determine this by making not one, but nine by-eye spline fits in LINER
to the emission line complex.  These nine fits will result in nine
different final values of each line parameter.  We take the mean of each
set of nine values as the true determination of that parameter.  The
scatter of the nine values around that mean is then used as the second
component in our uncertainty.  In all nine cases, the fit to the
pseudo-continuum is reasonable, but several were purposefully fit to the
outer envelope of the noise.  The two components of the uncertainty are
then added in quadrature to give the final value of the uncertainty
provided in this paper.

\subsection{Mrk\,1044 Emission Lines} \label{sec:mrkemi}

While this paper is focused around the absorption line results, we
measure and report the emission line measurements so as to place the
level of Mrk\,1044's metallicity in the broader context of most AGN
metallicity studies in which it is not feasible to measure the
absorption line properties.  The parameters for the observed emission
lines can be found in Table~\ref{tab:emission}.  Given for each line
(Lyman$\beta$, \ion{N}{3}, and \ion{O}{6}) are the equivalent widths, the
gaussian FWHM for any components they have, and the velocity offset from
systemic.  In general, the emission lines are offset by hundreds of
\,km\,s$^{-1}$ to the blue.  There does not appear to be a consistent
velocity offset among the emission lines.  The \ion{N}{3} doublet is
weak enough so as to be fit with a single gaussian each, while the
\ion{O}{6} doublet requires a narrow and broad component each.  The
Lyman$\beta$ line only requires a broad line; a narrow line does not
significantly improve the overall fit.  In general, gaussians
adequately, but not well fit the data.  As discussed above in
\S\ref{sec:analysis}, if left completely unconstrained the model of the
absorption lines returned very unphysical results with regards to the
doublets.  To ameliorate this, for each doublet the naturally weaker of
the lines is pinned to the stronger.  We pin the width and the
wavelength, but leave the relative fluxes unpinned because otherwise the
model cannot even adequately fit the data.  Due to the absence of a
narrow component to Lyman$\beta$ and because it is blended with
\ion{O}{6}, one may ask if it is properly identified.  Evidence in favor
of such an identification is: 1) a four-component fit to the
\ion{O}{6} complex (2 narrow, 2 broad) always leaves an excess of
blueward flux, 2) a six-component fit (3 per
\ion{O}{6}) is not sufficiently superior to a broad Ly$\beta$ +
four-component \ion{O}{6} and 3) the centroid of the component blueward
of the \ion{O}{6} is generally consistent with the redshift of the other
emission lines (Ly$\beta$, \ion{O}{3}).

\subsection{Mrk\,1044 Absorption Line Systems} \label{sec:mrkline}

We report good ($3\sigma$) detections of four lines belonging to
absorption systems intrinsic to Mrk\,1044.  We find a velocity for
System 1 of $-1158$\,km\,s$^{-1}$ and a velocity for System 2 of
$-286$\,km\,s$^{-1}$ relative to the systemic velocity of Mrk\,1044.
These lines are consistent with the systems found in F05, which reported
velocities of $-1158$ and $-286$\,km\,s$^{-1}$ (a perfect match!).  The
velocities of each individual line also match well with the mean
velocities.  In the case of System 2, the weaker of the \ion{O}{6} lines
is not detected at $>3\sigma$.  Table~\ref{tab:mrk1044} gives the
measured and calculated parameters for these absorption systems: the
observed wavelength, the FWHM, the equivalent widths, the calculated
column densities (by both the curve of growth and apparent optical depth
methods) and the velocity offset from the systemic.  For the lines that
are not resolved (System 2's) we give only the column density derived
from the curve of growth method.  For System 1 we also give the apparent
optical depth column density, though it should be noted that only the
\ion{O}{6} lines are definitely resolved ($>3\sigma$ above the
$2\,\Delta\lambda_{lsf}$ limit which is about 40\,km\,s$^{-1}$).
Additionally, the $3\sigma$ upper limits on the undetected \ion{N}{3}
lines are given.

To determine the column densities with the apparent optical depth method, we first calculate the covering fraction ($C_f$) of each of these two systems following \citet{H97}.  However, there is a problem in that \ion{O}{6}\,$\lambda1038$ (I$_1$ in their formalism) has a lower value of the normalized flux in its trough than \ion{O}{6}\,$\lambda1032$ (I$_2$), on average about 0.10 to 0.15, respectively.  This is unphysical, and we attempt to solve this problem by modifying the lines as follows.  We note that the errors on points in these two troughs are about 0.05, and we calculate two covering fractions, one where the trough of I$_2$ is lowered by one sigma, and one where the trough of I$_1$ is raised by one sigma.  This allows seemingly reasonable values of the covering fraction to be found, typically around 0.91 and 0.86 for the two methods.  Unfortunately, the covering fractions are often barely too small in the majority of cases where I$_2$ is lowered.  We report the values derived from increasing the trough of I$_1$ by one sigma as our values of \ion{O}{6}.  In the case of \ion{O}{6}\,$\lambda1032$ the calculation of the optical depth just fails ($C_f+I_{trough}=1$).  For this line we average over the seven calculable results.  While this line has an uncertainty on its column density even greater than its value, it should be noted that it matches the value from the other line well.  The covering factor for System 2 is, by the formalism, set to 1 since I$_2<$I$_1^2$.  While this is also unphysical, such a result is not unexpected due to this system being unresolved, let alone minimally detected.  This result merely tells us that we have no information about the covering fraction for this line.

The value of the covering fraction for System 1's \ion{O}{6} is different than that found for \ion{C}{4} and \ion{N}{5}.  As just given above, we find a value of about 0.86 for \ion{O}{6}, while the derived values for the other lines are approximately 0.72.  One solution is that the location in the broad-line region (BLR) that the \ion{O}{6} emission line flux comes from is sufficiently closer to the central black hole than the location that produces the \ion{C}{4} and \ion{N}{5}.  The absorbing gas (System 1), being of finite size, will likely completely cover the continuum source, mostly cover the \ion{O}{6} emitting region, and cover to a lesser degree the \ion{C}{4} and \ion{N}{5} emitting region.  Such a toy model could explain the observed covering fraction for these three species.  For this reason we do not consider this discrepancy in covering fraction a problem to be corrected.

There is one major caveat that must be addressed with regards to the measurement of the Ly$\beta$ line of System 1, namely that this line is heavily blended with a galactic H$_2$ line.  Given the difficulty in subtracting the H$_2$ (see \S\ref{sec:analysis}), one can see how difficult getting an accurate value of the \ion{H}{1} column density is.  The uncertainties in the measurements of its properties are very large compared to the other absorption lines, reflecting this.  What isn't reflected in those uncertainties is any systematic effect.  Comparing to the Ly$\alpha$ derived values from F05, we find that the Ly$\beta$ value of \ion{H}{1} is much higher, by a factor of 2-4.  If this is so, the H$_2$ at this location is under-subtracted, which matches the effect on the strong H$_2$ lines.  In this case, our Ly$\beta$ derived value should represent more of an upper limit on the total \ion{H}{1} column.  Because which \ion{H}{1} value we decide to take will affect our metallicity measurements (in magnitude, but not in direction or in general effect) later in \S\ref{sec:conclude}, we wish to point out that there is no reason to trust the Ly$\beta$ column above that of the Ly$\alpha$.  As Figure 4 of F05 shows, the Ly$\alpha$ line is symmetric and appears clean of contamination from coincident lines.  Its profile does not suggest saturation and the continuum around is well-normalized.

To determine the elemental column densities from the ionic column densities, we must make an ionization correction.  To do so, we use the photoionization-equilibrium code Cloudy\,94\footnote{Cloudy version C94.00, obtained from the Cloudy webpage http://www.nublado.org/} \citep{cloudy}.  This code creates a model of gas in equilibrium with an incident flux.  From this, we take the predicted column densities of various species for specific input conditions and compare them with the data.  The four inputs to Cloudy we are concerned with are 1) the incident spectral energy distribution (SED), 2) $U$, the ratio of the number densities of ionizing photons to particle (H) density at the surface of the modeled cloud, 3) the abundance ratios in the gas cloud, and 4) $N_H$, the column density of hydrogen through the cloud.  Throughout our analysis we keep the SED and the assumed metallicity the same and vary $U$ and $N_H$.  For the SED we focus on Cloudy's standard AGN template for much of our analysis, and for abundances we select solar in level and in mixture also for most of our analysis.  We then calculate the column densities of various ions at a grid of 0.01 spacing in both $\log U$ from $-2$ to 0 and $\log N_{H}$ from 18 to 20.

We then look for models which predict column densities in agreement with all those measured.  In this study we did find the column density of \ion{O}{6}.  We are not, however, able to determine the physical parameters of this absorber with just this single species.  In addition to this information found by FUSE observations, we bring into this study the column densities of \ion{H}{1}, \ion{C}{4}, and \ion{N}{5} found by using STIS in F05\footnote{The values of the covering fraction in that paper had an error, resulting in low column densities of \ion{C}{4} and \ion{N}{5}.  The correct mean values as reported in the erratum are also quoted here.}  The column densities of \ion{C}{4} and \ion{N}{5} are $14.47\pm0.06$ (\ion{C}{4}) and $14.46\pm0.02$ (\ion{N}{5}).  Because Ly$\alpha$ is not a doublet, its covering fraction cannot be truly determined.  Since we determine different covering fractions for multiple doublets, we cannot just assign it a covering fraction determined from other lines.  Instead, we find the column density of \ion{H}{1} for a range of covering fraction from 0.75 to 1.00.  The results are shown in Figure~\ref{fig:unh}.  In this figure the models which match the observed column densities to within one sigma are shaded for each ion.  One can see that there is a small region of parameter space around $\log U=-1.29$, $\log N_H=18.85$ where all three of the metals agree to within one sigma.  There is no location, however, in which the three metal lines and the hydrogen agree.

We also investigate how the choice of incident SED affects our results.  We use SEDs of two NLS1s (Ark\,564 and NGC\,4051) and find qualitatively the same result, though the agreement of the three metal lines takes place at the 3-3.5$\sigma$ level.  Compared to Figure~\ref{fig:unh}, all metals shift to lower $\log U$ values, the magnitude of which is highest for Oxygen, small for Nitrogen, and smallest for Carbon, about 1.0, 0.5, and 0.3 dex respectively.  Additionally, the amount of \ion{H}{1} for a particular $\log U$ shifts to lower $N_H$, between 0.4 and 0.6 dex.  We also decrease the amount of flux coming out in the EUV (an unobservable region of the spectrum) to test its effects and find that the magnitude of its changes is small compared to the SED differences between the standard ``table agn'' and the the two NLS1s.  We also test non-solar abundance mixtures by increasing the nitrogen by factors of two and four.  We find at the $3\sigma$ level our data is consistent with overabundant Nitrogen at twice the solar mixture, but inconsistent with four-times-solar.

\subsection{Galactic Absorption Lines} \label{sec:galline}

The interstellar medium imprints absorption lines belonging to molecular
hydrogen, \ion{C}{2}, \ion{O}{1}, \ion{Ar}{1}, and \ion{Fe}{2}.  The
measured parameters of these lines are given in
Table~\ref{tab:galactic}.  Given are the ions, their rest and observed
wavelengths, the FWHM, equivalent width, column densities via the two
methods, and the relative velocity (to zero).  We also give the value
for the H$_2$ column density as fit to the weak absorption lines.  The
two \ion{C}{2} lines were blended to some degree with H$_2$, \ion{C}{2}*
especially, and the fit was unstable in SPECFIT.  Additionally, the
values of the column density between the two carbon lines are
inconsistent for all values of $b$.  For that reason, only their observed parameters (EW, FWHM, $v_{rel}$) are given in the table.  We are not
able to give column density values for the two \ion{Fe}{2} lines as
their f-values are not available (curve of growth method) and neither
are definitely resolved (apparent optical depth method).  In addition,
the redward Fe line ($\lambda1063$\AA) is blended in its wings with
H$_2$.  This leaves us with only \ion{O}{1} and \ion{Ar}{1} with
possible column density values.  The \ion{O}{1} line appears to be
definitely resolved, so we give its column density via the apparent
optical depth method.  Unfortunately, this leaves us without multiple
lines from a single species, which means we are unable to determine the
correct $b$ parameter and thus we cannot find the column density via the
curve of growth.  Calculating the column density for \ion{O}{1} for
multiple $b$ values gives a best match to the apparent optical depth
column density at $b$ between 20 and 40\,km\,s$^{-1}$.  Unfortunately, \ion{O}{1}$\lambda1032$ as found in F05 has contradictory parameters.  As mentioned in that study, \ion{O}{1}$\lambda1032$ is found in the low-resolution G140L spectrum and is possibly a blend.  For this reason, there is no $b$ value consistent with the \ion{O}{1} from both studies.  With this in mind, we
simply return column densities for a $b$ value of 20\,km\,s$^{-1}$ and
remind the reader about possible accuracy issues.  One should also note that the three clean lines (\ion{O}{1}, \ion{Ar}{1} \& \ion{Fe}{2}1055) have a velocity consistent with zero, while the three other lines have the most discrepant velocities and are also the lines blended with H$_2$.

\subsection{Intergalactic Absorption} \label{sec:igm}

On account of the findings in F05, we search our data for the presence of intergalactic absorption lines.  In that paper three Ly$\alpha$ forest lines were discovered (see their Table 5).  Assuming no saturation (still on the linear part of the curve of growth), the expected equivalent width of the strongest Ly$\beta$ line would be about 40\,m\AA.  The three sigma detection in this region is around 100\,m\AA.  Regardless, there are two absorption features that may be the Ly$\beta$ counterparts of the two stronger of the Ly$\alpha$ absorption systems.  The weakest of the three Ly$\alpha$ systems lies coincident with the blend of Galactic H$_2$ and \ion{C}{2} lines and given the problems with deblending described above in \S\ref{sec:analysis} along with its just-three-sigma detection in F05, the recovery of it should be unreasonable.  The other two Ly$\alpha$ systems should have corresponding absorption in Ly$\beta$ at 1033.0 and 1035.8\AA, and we find two systems at 1033.1 and 1035.7\AA\ with equivalent widths of 105 and 80\,m\AA.  By this measure, the weaker line is not definitely detected and the stronger is very close to the nominal three sigma limit.  It should be noted that at the location of the stronger line the pseudo-continuum is at an apparent minimum, lying between the \ion{O}{6} emission line and some excess flux that rises towards the blue for about 5\AA (see Figure~\ref{fig:fullspec}).  The normalization of this region is very uncertain and contributes a 20\,m\AA\ uncertainty to the overall equivalent width error budget.  Combined with the photon noise error, this pushes the detection of the stronger of the lines below the three sigma limit.  While both of these features are in the correct locations to be the Ly$\beta$ counterparts to the low-redshift Ly$\alpha$ forest lines found in F05, the quality of the spectrum at this location is not sufficient to state a definitive confirmation.  Were these the Ly$\beta$ lines, their strengths would be much larger than expected, but still consistent at the two sigma level (indeed consistent with zero at the three sigma level).  For this reason we suggest the Ly$\alpha$ determined values for this system stand as the measurement of these systems.

\section{Discussion and Conclusions} \label{sec:conclude}

\begin{figure}
\epsscale{0.9}
\plotone{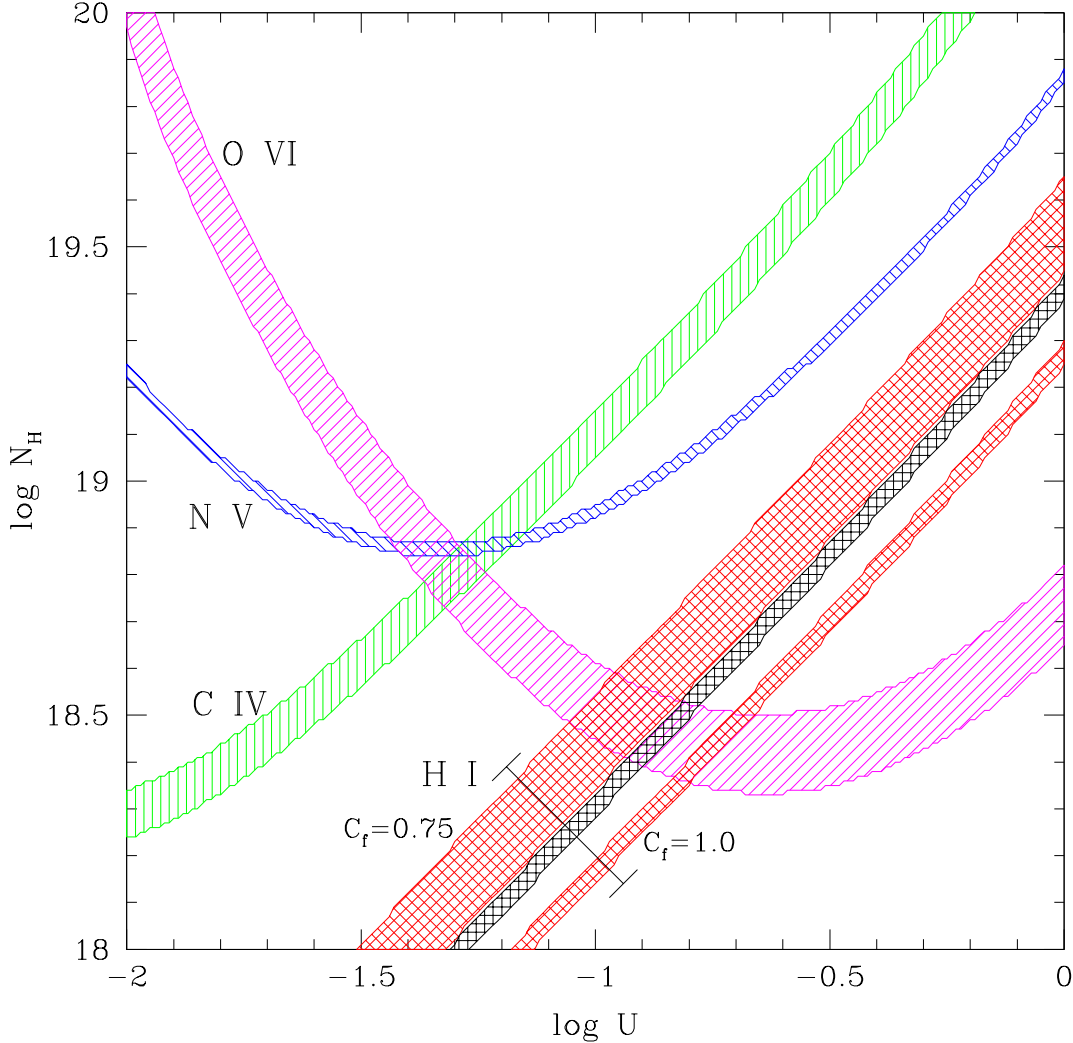}
\caption{\label{fig:unh}
CLOUDY models with its AGN template spectrum at many $\log U$-$\log N_H$
points assuming solar metallicity.  Shaded regions indicate agreement
with observed column densities of several ions at the $1\,\sigma$ level.
Slashed shading represents agreement with the observed \ion{O}{6} column
density, vertical shading indicates agreement with \ion{C}{4},
backslashed with \ion{N}{5}, and crosshatched with \ion{H}{1}.  Because
the covering fraction ($C_f$) parameter differs between \ion{O}{6} and
both \ion{C}{4}, and \ion{N}{5}, column densities of \ion{H}{1} are
calculated for several $C_f$.  Shown are those for $C_f=0.75$, 0.85 and
1.00, with the lower covering fraction preferring a higher $N_H$.  The
range of models in agreement (the thickness of the band) with \ion{H}{1}
is relatively constant down to about $C_f=0.80$.  Below this, the
uncertainty in the column density rises dramatically because the depth
of the Ly$\alpha$ line is $\approx0.75$ in terms of normalized flux.
See Fig. 4 of F05 for an example of the normalized spectrum around
Ly$\alpha$.  The metal lines all agree in a small region of parameter
space around $\log U=-1.29$, $\log N_H=18.85$.  The vertical distance
between this point and the preferred model for \ion{H}{1} is about
$+0.7$, between the metal-selected model and just the $1\,\sigma$
envelope of \ion{H}{1} is about $+0.6$.  }\end{figure}

As shown in \S\ref{sec:mrkline}, there is no one set of input conditions from which Cloudy can create a model that reproduces all of the observed column densities of the associated absorption lines.  This indicates that one or more assumptions made in the creation of those models must be incorrect.  The fact that there is a model ($\log U=-1.29, \log N_H=18.85$ with the standard Cloudy AGN SED) which is extremely satisfactory in predicting the column densities of all the metal species, but not that of \ion{H}{1} provides one answer.  If the bulk metallicity is in fact about 0.7 dex higher (i.e. $N_H$ really 18.15), then all measured lines (assuming Ly$\alpha$ has $C_f\approx0.75$) are in excellent ($<1\sigma$) agreement.  The fact that there is one point that all three metal species agree so well at indicates a good probability of having a solar mixture.  With a bulk metallicity around five times solar ($+0.7$ dex) this is inconsistent with the concept of N/O scaling like O/H \citep[and references therein]{HF}, at least for this object.  If nitrogen did scale with metallicity, then, under this solution, the \ion{N}{5} curve in Figure~\ref{fig:unh} should like as far above the \ion{C}{4}-\ion{O}{6} agreement point as the \ion{H}{1} lies below it, which it clearly does not.  The best-fit Nitrogen is at most 0.06 dex above such a point.  At most, Nitrogen can be overabundant with respect to the solar mixture at the factor of two level, above which it shares no model in common with both Carbon and Oxygen.  Understandably then, this paper gives a much different value of $Z$ than that found in F05 (minimum$\approx1.5$ solar) where it was assumed that $N\propto Z^2$.

\begin{figure}
\epsscale{0.8}
\plotone{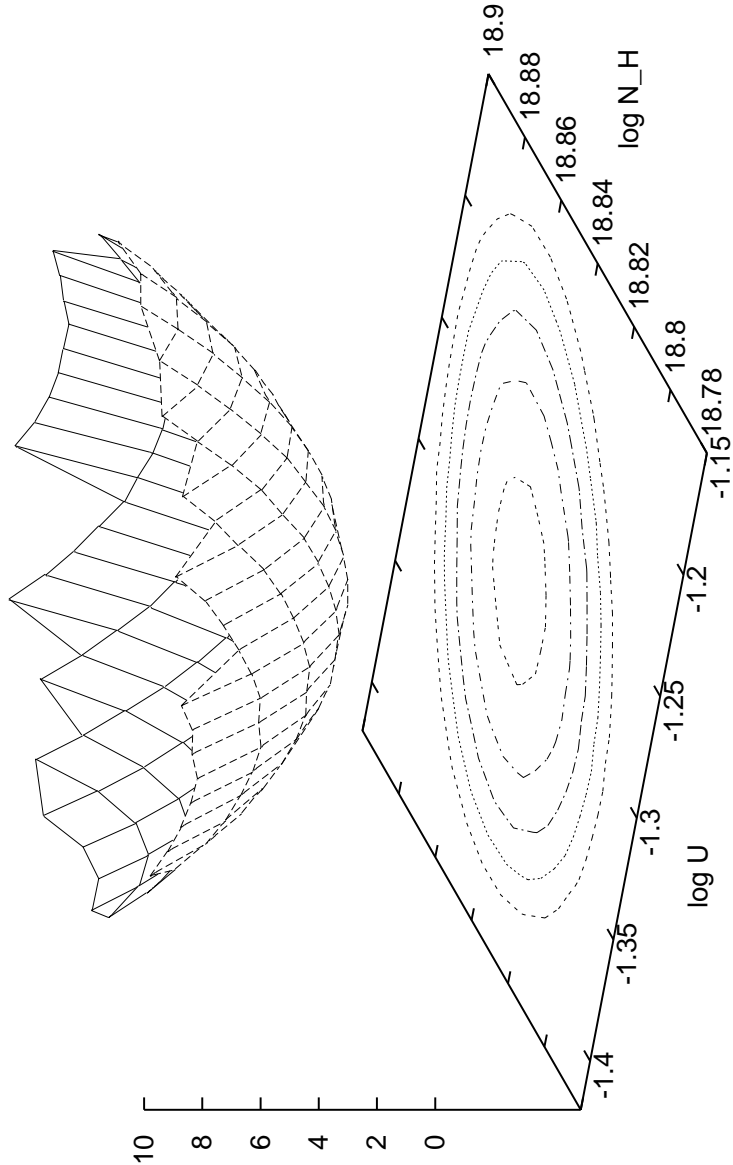}
\caption{\label{fig:solar}
The $\chi^2$ surface for Cloudy models at solar metallicity compared to
observed column densities of \ion{H}{1}, \ion{C}{4}, \ion{N}{5}, and
\ion{O}{6}.  The surface is shown up to a $\chi^2$ of 10, and the curves
of constant $\chi^2$ (2, 4, 6, 8, and 10) are projected into the $\log
U$-$\log N_H$ plane.  The \ion{H}{1} component of the fit has been
shifted by $-0.71$ dex to match the metal's minimum at $\log U=-1.29$,
$\log N_H=18.85$.  This implies a metallicity of $+0.71$ and thus a
corrected $\log N_H$ of 18.14.}\end{figure}

\begin{figure}
\epsscale{0.8}
\plotone{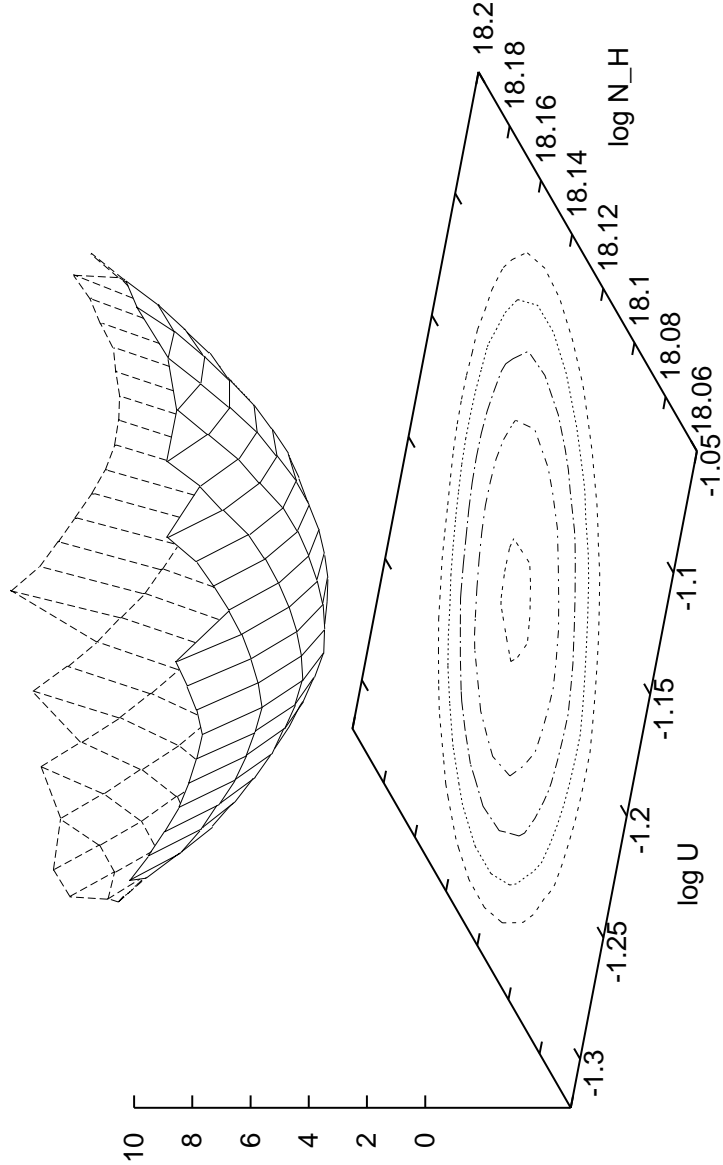}
\caption{\label{fig:five}
The $\chi^2$ surface for Cloudy models at a bulk metallicity of five
times solar with a solar mixture in metals and Helium enhanced by $\Delta Y/\Delta Z=2$ compared to observed column
densities of \ion{H}{1}, \ion{C}{4}, \ion{N}{5}, and \ion{O}{6}.  The
surface is shown up to a $\chi^2$ of 10, and the curves of constant
$\chi^2$ (2, 4, 6, 8, and 10) are projected into the $\log U$-$\log N_H$
plane.  The minimum lies at $\log U=-1.20$, $\log N_H=18.13$, only
slightly different than the inferred values from
Figure~\ref{fig:solar}.}\end{figure}

The effects on the metallicity by the SED chosen can be large.  As mentioned in \S\ref{sec:mrkline}, the metal lines shift to lower values of $\log U$ when a NLS1-specific SED is used, the result of which is that the agreement point is around $\log U=-1.7$.  The major change, however, is the shift of \ion{H}{1} to lower values of $\log N_H$ which does raise the inferred value of the metallicity by another factor of 2-4 (between 10 and 20 times solar) with the SED of NGC\,4051 providing the highest metallicity.  To confirm these extremely high values of the metallicity we rerun Cloudy with the metals set to five or ten times the solar value for the standard and NLS1 SEDs respectively.  To be fully self-consistent, we also take Helium enrichment into account.  For one model of Helium enrichment ($\Delta Y/\Delta Z=2$) and bulk metallicity around ten times solar, $\log(He/H)\approx-0.52$.  Including this effect does not change our inferred metallicities, but it does shift the best-match $\log U$ to slightly higher values.  At the increased metallicities, the metal lines are put into agreement with the \ion{H}{1}, though this point is also at larger values of $\log U$ (cumulative with He effects).  For the standard AGN SED, we find $\log U=-1.20\pm0.04$, $\log N_H=18.13\pm0.02$, and $Z=5^{+2}_{-1}$ solar metallicities.  For the Ark\,564 SEDs we find $\log U=-1.55\pm0.03$, $\log N_H=17.86\pm0.02$, and $Z=17^{+5}_{-3}$.  For the NGC\,4051 SED we find $\log U=-1.68\pm0.03$, $\log N_H=17.85\pm0.02$, and $Z=22^{+7}_{-4}$.  The $\chi^2$ surface for the standard AGN SED with solar metallicity is shown in Figure~\ref{fig:solar} and for the standard AGN SED with five times solar metallicity in Figure~\ref{fig:five}.  The lines of constant $\chi^2$ are projected into the $\log U$-$\log N_H$ plane.  The smoothness and parabolic shape of the $\chi^2$ surface indicate that we have well-sampled the parameter space.

For completeness' sake, we remind the reader that we have used the Ly$\alpha$-derived value of \ion{H}{1} which is much smaller than the Ly$\beta$-derived value.  Even assuming that the Ly$\beta$ value is the correct one (which we conclude should not be done), the inferred bulk metallicity is $+0.3$ dex (twice solar) using the same arguments.

This study also reinforces that one cannot confine oneself to a small section of the electromagnetic spectrum if one wants to accurately model the physical conditions in these absorbing systems.  If one created the same models as we did in this study, but only used the F05 data and assumed $N\propto Z^2$, one would conclude $\log U\approx-1.8$, as that is where the \ion{N}{5} and \ion{H}{1} are equidistant from \ion{C}{4}.  While one would infer a metallicity value approximately $+0.7$ dex, the physical conditions would be very different and the photoionization correction for other species could be quite wrong.  It is only because we have FUSE and HST data that we are able to determine the correct physical conditions and thus the abundance level and mixture.

With the results from the absorption lines, we can finally compare this to the values expected of the emission lines and check for agreement.  We compare to the line ratios formalism given in \citet{H02}.  We note that the solar mixture used in \citet{H02} is with the earlier, more metal rich values.  If the new values for the solar metallicity are to be used, the metallicities referenced here should be considered about -0.11 less as per \citet{B03}.  The emission line ratios in common (using this study's and F05's corrected values) are \ion{N}{5}/\ion{He}{2}, \ion{N}{5}/\ion{C}{4}, \ion{N}{5}/\ion{O}{6}, and \ion{N}{5}/(\ion{C}{4}$+$\ion{O}{6}).  We also have a measurement of \ion{N}{4}], but \citet{H02} find the metallicities derived from it to be discrepant, as do we.  We therefore exclude it from this comparison.  Ratios involving \ion{N}{5} do surprisingly well, indicating a metal rich gas of $+0.7$ to $+0.8$ on average, very close to our absorption line values using the standard AGN SED.  This is curious, however, in that these emission line models assume $N\propto Z^2$ which is very inconsistent with the results of this absorption line study, and yet get the same value of the bulk metallicity.  In other words, emission line ratios based off of \ion{N}{5} (such as \ion{N}{5}/\ion{C}{4}) appear to be good indicators of $Z$, but perhaps for the wrong reason for this particular system under investigation.  Compared to the NLS1 SEDs, however, the emission line ratios fare rather poorly.

The results of this abundance study are somewhat surprising, not because of the super-solar bulk metallicity found (which was already expected), but because of the solar mixture, especially nitrogen relative to carbon and oxygen.  Theoretical models and observations around solar metallicity have nitrogen scaling like $Z^2$ ([N/O] $\propto$ [O/H]) because the CNO cycle will preferentially convert oxygen and carbon into $^{14}$N, enhancing nitrogen relative to the other metals over long timescales \citep[and references therein]{HF}.  This theory then goes against our expectations for Mrk\,1044.  One would assume that a local spiral galaxy such as Mrk\,1044 would have had constant star formation over the past age of the universe, and therefore have nitrogen scaling as $Z^2$.  There exists, then, a contradiction between our data and theory.  One way to reconcile this would have some special enrichment process occur in the nucleus of Mrk\,1044 which would enhance the metallicity level and mixture to that which we observe.  Another is that the existing models are simply not appropriate to this system.  The theory which predicts N going like $Z^2$ results from studies near solar metallicity.  It is fair to say that a metallicity of $+0.7$ to $+1.0$ (such as in Mrk\,1044) is a far extrapolation from the data these trends are based upon.  Additionally, we find that metal mixtures such as we find are not unprecedented for high metallicity stars.  A recent study of planet-bearing stars (i.e. metal-rich stars) by \citet{planet} finds that [N/H] scales with [Fe/H] ( [N/H]/[Fe/H] slope is consistent with zero at the two sigma level) over the range $-0.4<$[Fe/H]$<+0.4$.  Said study stops short of the significantly super-solar status of this Seyfert since the set of such stars simply subsides.
Thus our result can simply be a continuation of an existing observed trend in enrichment.  This, like the $N\propto Z^2$ theory, hinges on an extrapolation, though not nearly as large a one.  With the dearth of Galactic studies at extremely high metallicities, Mrk\,1044 can provide a calibration point not only for AGN metallicity studies, but also for enrichment theory.

\begin{acknowledgements}

The authors wish to thank Marc Pinsonneault for discussions involving
metal enrichment and star formation histories.  This research has made
use of the NASA/IPAC Extragalactic Database (NED) which is operated by
the Jet Propulsion Laboratory, California Institute of Technology, under
contract with the National Aeronautics and Space Administration.
Primary support for this work was provided by NASA grant NNG04GI04G.

\end{acknowledgements}

\clearpage

%\newpage
\begin{deluxetable}{ccccc}
%\rotate
\tablecaption{Emission Line Properties of Mrk\,1044
\label{tab:emission}}
\tablehead{
\colhead{Ion}
&\colhead{$\lambda_{Rest}$}
&\colhead{Equivalent}
&\colhead{FWHM}
&\colhead{Velocity}\\
\colhead{}
&\colhead{[\AA]}
&\colhead{Width [\AA]}
&\colhead{[km\,s$^{-1}$]}
&\colhead{[km\,s$^{-1}$]}}
\startdata
\ion{N}{3}&$990$&$0.68\pm0.15$&$341\pm54$&$-190\pm8$ \\
\ion{N}{3}&$992$&$0.55\pm0.14$&& \\
\ion{H}{1}&$1026$&$5.66\pm3.65$&$3400\pm1000$&$-314\pm8$ \\
\ion{O}{6}&$1302$\,Narrow&$3.78\pm0.53$&$656\pm45$&$-494\pm8$ \\
\ion{O}{6}&$1302$\,Broad&$25.7\pm5.1$&$3720\pm480$&$-880\pm8$ \\
\ion{O}{6}&$1308$\,Narrow&$6.10\pm0.60$&& \\
\ion{O}{6}&$1308$\,Broad&$13.1\pm4.8$&& \\
\enddata
\end{deluxetable}

%\newpage
\begin{deluxetable}{ccccccc}
%\rotate
\tablecaption{Measured and Calculated Parameters of Mrk\,1044
\label{tab:mrk1044}}
\tablehead{
\colhead{Line}
&\colhead{Observed}
&\colhead{FWHM}
&\colhead{Equivalent}
&\colhead{log(Column)}
&\colhead{log(Column)}
&\colhead{Velocity}\\
\colhead{}
&\colhead{Wavelength [\AA]}
&\colhead{[km\,s$^{-1}$]}
&\colhead{Width [m\AA]}
&\colhead{[$cm^{-2}$]\,\tablenotemark{a}} 
&\colhead{[$cm^{-2}$]\,\tablenotemark{b}}
&\colhead{[km\,s$^{-1}$]}}
\startdata
System 1&&&&& \\
Ly$\beta$&$1038.5757$&$50\pm12$&$135\pm37$&$14.69$&$14.60$&$-1175\pm8$ \\
%Ly$\beta$&$1038.5757$&$50\pm12$&$135\pm37$&$14.69^{+0.23}_{-0.23}$&$14.60^{+0.09}_{-0.11}$&$-1175\pm7$ \\
\ion{O}{6}$1032$&$1044.9335$&$69\pm3$&$244\pm14$&$15.12^{+0.16}_{-0.14}$&$14.93^{+0.49}_{-14.93}$&$-1153\pm8$ \\
\ion{O}{6}$1038$&$1050.7038$&$61\pm3$&$210\pm15$&$15.08^{+0.14}_{-0.12}$&$14.92^{+0.08}_{-0.10}$&$-1150\pm8$ \\
\ion{N}{3}$990$\,\tablenotemark{c}&$1002.291$&&&$<14.12$&& \\
\ion{N}{3}$992$\,\tablenotemark{c}&$1004.103$&&&$<14.16$&& \\
Ly$\alpha$\,\tablenotemark{d,e}&$1230.9819$&$92\pm7$&$341\pm5$&&$14.20^{+0.09}_{-0.12}$&$-1156\pm7$ \\
\ion{N}{5}$1239$\,\tablenotemark{d}&$1254.4633$&$71\pm3$&$209\pm8$&&$14.42^{+0.02}_{-0.02}$&$-1143\pm6$ \\
\ion{N}{5}$1243$\,\tablenotemark{d}&$1258.4999$&$71\pm3$&$162\pm6$&&$14.51^{+0.02}_{-0.03}$&$-1143\pm6$ \\
\ion{C}{4}$1549$\,\tablenotemark{d}&$1567.7329$&$50\pm6$&$204\pm5$&$13.94^{+0.02}_{-0.03}$&$14.26^{+0.23}_{-0.53}$&$-1147\pm5$ \\
\ion{C}{4}$1551$\,\tablenotemark{d}&$1570.3602$&$47\pm5$&$156\pm4$&$14.04^{+0.02}_{-0.02}$&$14.48^{+0.05}_{-0.06}$&$-1143\pm5$ \\
System 2&&&&& \\
\ion{O}{6}$1032$&$1047.8857$&$24\pm8$&$41\pm10$&$13.57^{+0.10}_{-0.13}$&&$-295\pm8$ \\
\ion{O}{6}$1038$&$1053.7330$&$16\pm13$&$18\pm16$&$13.48^{+0.29}_{-0.89}$&&$-276\pm8$ \\
\ion{N}{3}$990$\,\tablenotemark{c}&$1005.102$&&&$<14.12$&& \\
\ion{N}{3}$992$\,\tablenotemark{c}&$1006.919$&&&$<14.16$&& \\
\enddata
\tablenotetext{a}
{Derived from curve of growth arguments}
\tablenotetext{b}
{Derived from optical depth integration}
\tablenotetext{c}
{$3\sigma$ upper limits}
\tablenotetext{d}
{Values from F05, corrected for error in $C_f$}
\tablenotetext{e}
{$C_f=0.75$}
\end{deluxetable}

%\newpage
\begin{deluxetable}{ccccccc}
%\rotate
\tablecaption{Galactic Absorption Features
\label{tab:galactic}}
\tablehead{
\colhead{Line}
&\colhead{Observed}
&\colhead{FWHM}
&\colhead{Equivalent}
&\colhead{log(Column)}
&\colhead{log(Column)}
&\colhead{Velocity}\\
\colhead{}
&\colhead{Wavelength [\AA]}
&\colhead{[\,km\,s$^{-1}$]}
&\colhead{Width [m\AA]}
&\colhead{[$cm^{-2}$]\,\tablenotemark{a}} 
&\colhead{[$cm^{-2}$]\,\tablenotemark{b}}
&\colhead{[\,km\,s$^{-1}$]}}
\startdata
\ion{C}{2}1036.337&1036.2431&$83.9\pm10.0$&$263\pm33$&&&$-27\pm8$ \\
\ion{C}{2}*1037.018&1036.8348&$236\pm53$&$610\pm180$&&&$-53\pm8$ \\
\ion{O}{1}1039.230&1039.2248&$53.5\pm4.2$&$183\pm13$&$15.75^{+0.10}_{-0.09}$&$15.56^{+0.05}_{-0.06}$&$-2\pm8$ \\
\ion{Ar}{1}1048.220&1048.2091&$17.8\pm4.5$&$93\pm14$&$13.73^{+0.09}_{-0.10}$&&$-3\pm8$ \\
\ion{Fe}{2}1055.262&1055.2443&$25.0\pm4.8$&$58\pm12$&&&$-5\pm8$ \\
\ion{Fe}{2}1063.2&1063.1610&$72\pm17$&$213\pm41$&&&$-11\pm8$ \\
%\ion{C}{2}1036.337&1036.2431&$83.9\pm10.0$&$263\pm33$&&&$-27.163\pm$ \\
%\ion{C}{2}*1037.018&1036.8348&$236\pm53$&$610\pm180$&&&$-52.961\pm$ \\
%\ion{O}{1}1039.230&1039.2248&$53.5\pm4.2$&$183\pm13$&$15.75^{+0.10}_{-0.09}$&$15.56^{+0.05}_{-0.06}$&$-1.500\pm$ \\
%\ion{Ar}{1}1048.220&1048.2091&$17.8\pm4.5$&$93\pm14$&$13.73^{+0.09}_{-0.10}$&&$-3.117\pm$ \\
%\ion{Fe}{2}1055.262&1055.2443&$25.0\pm4.8$&$58\pm12$&&&$-5.028.\pm$ \\
%\ion{Fe}{2}1063.2&1063.1610&$72\pm17$&$213\pm41$&&&$-10.997.\pm$ \\
H$_2$&&&&$\approx16.9$\,\tablenotemark{c}&&$\approx0$ \\
\enddata
\tablenotetext{a}
{Derived from curve of growth arguments}
\tablenotetext{b}
{Derived from optical depth integration}
\tablenotetext{c}
{By-eye choices fluctuated between (7-8)$\times10^{16}$}
\end{deluxetable}

%\newpage
%\clearpage

%\newpage
%\clearpage

%\newpage
%\clearpage

%\newpage
%\clearpage

%\newpage
%\clearpage

%\newpage
%\clearpage


\begin{thebibliography}{}

\bibitem[Arav et al.(2005)]{A05}
Arav, N., Kaastra, J., Kriss, G.A., Korista, K.T., Gabel, J., Proga, D.  2005, \apj, 620, 655

\bibitem[Bahcall \& Kozlovsky(1969)]{BK69}
Bahcall, J. \& Kozlovsky, B.-Z.  1969, \apj, 155, 1077

\bibitem[Baldwin et al.(2003)]{B03}
Baldwin, J.A., Hamann, F., Korista, K.T., Ferland, G.J., Dietrich, M., \& Warner, C.  2003, \apj, 583, 649

\bibitem[Boller, Brandt \& Fink(1996)]{BBF}
Boller, Th., Brandt, N., \& Fink, H.  1996, A\&A, 305, 53

\bibitem[Boroson(2002)]{B02}
Boroson, T.  2002, \apj, 565, 78

\bibitem[Boroson \& Green(1992)]{BG}
Boroson, T. \& Green, R.  1992, \apjs, 80, 109

\bibitem[Brandt, Mathur \& Elvis(1997)]{BMF}
Brandt, N., Mathur, S. \& Elvis, M.  1997, \mnras, 285L, 25

\bibitem[Ecuvillon et al.(2004)]{planet}
Ecuvillon, A., Israelian, G., Santos, N.C., Mayor, M., Garcia Lopez, R.C., Randich, S.  2004, A\&A, 418, 703

\bibitem[Ferland et al.(1998)]{cloudy}
Ferland, G. J., Korista, K. T., Verner, D. A., Ferguson, J. W., Kingdon, J. B., Verner, E. M.  1998, PASP, 110, 761.

\bibitem[Fields et al.(2005)]{hst}
Fields, D., Mathur, S., Pogge, R.W., Nicastro, F., \& Komossa, S.  2005, \apj, 620, 183

\bibitem[Ferrarese \& Merritt(2000)]{FM00}
Ferrarese, L., \& Merritt, D. 2000, ApJ, 539, L9

\bibitem[Gebhardt et al.(2000)]{G00}
Gebhardt, K., et al. 2000a, ApJ, 539, L13

\bibitem[Grupe \& Mathur(2004)]{GM04}
Grupe, D. \& Mathur, S. 2004, \apj, 606L, 41

\bibitem[Grupe et al.(2004)]{G04}
Grupe, D., Wills, B.J., Leighly, K.M., \& Meusinger, H.  2004 \aj, 127, 156

\bibitem[Grupe et al.(2005)]{G05}
Grupe, D., Mathur, S., Wills, B.J., \& Osmer, P.  2005 \aj, submitted

\bibitem[Hamann et al.(1997)]{H97}
Hamann, F., Barlow, T.A., Junkkarinen, V., Burbidge, E.M.  1997, \apj, 478, 80

\bibitem[Hamann \& Ferland(1999)]{HF}
Hamann, F. \& Ferland, G.  1991, \araa, 37, 487

\bibitem[Hamann et al.(2002)]{H02}
Hamann, F., Korista, K.T., Ferland, G.J., Warner, C., \& Baldwin, J.  2002, \apj, 564, 592

\bibitem[Kriss(1994)]{SPECFIT}
Kriss, G.A.  1994 in Astronomical Data Analysis Software \& Systems III, A.S.P. Conf. Series, Vol. 61, ed. D. R. Crabtree, R. J. Hanisch, \& J. Barnes (Astronomical Society of the Pacific: San Francisco), p. 437.

\bibitem[Mathur(2000)]{M2K}
Mathur, S.  2000, \mnras, 314, L17

\bibitem[Mathur et al.(2001)]{M01}
Mathur, S., Matt, G., Green, P.J., Elvis, M., Singh, K.P.  2001, \apj, 551L, 13

\bibitem[Mathur \& Grupe(2005)]{MG05}
Mathur, S. \& Grupe, D. 2005, A\&A, in press

\bibitem[Merritt \& Ferrarese(2001)]{MF01}
Merritt, D., \& Ferrarese, L. 2001, ApJ, 547, 140

\bibitem[Osterbrock \& Pogge(1985)]{OP}
Osterbrock, D.E. \& Pogge, R.W.  1985, \apj, 297, 166

\bibitem[Pogge \& Owen(1993)]{LINER}
Pogge, R.W., \& Owen, J.M. 1993, LINER; An Interactive Spectral Line Analysis Program, OSU Internal Report 93-01

\bibitem[Pounds, Done \& Osborne(1995)]{P95}
Pounds, K., Done, C., \& Osborne, J.  1995, \mnras, 277, L5

\bibitem[Romano et al.(2002)]{ark564}
Romano, P., Mathur, S., Pogge, R.W., Peterson, B.M. \& Kuraszkiewicz, J.  2002, \apj, 578, 64

\bibitem[Romano et al.(2004)]{arksed}
Romano, P., Mathur, S., Turner, T.J., Kraemer, S.B., Crenshaw, D.M., Peterson, B.M., Pogge, R.W., Brandt, W.N., George, I.M., Horne, K., Kriss, G.A., Netzer, H., Shemmer, O., Wamsteker, W.  2004, \apj, 602 635

\bibitem[Savage \& Sembach(1991)]{SS91}
Savage, B.D. \& Sembach, K.R.  1991, \apj, 379, 245

\bibitem[Sembach(1999)]{galline}
Sembach, K.  1999 in Stromlo Workshop on High-Velocity Clouds, A.S.P. Conf. Series, Vol. 166, eds. B.K. Gibson \& M.E. Putman p. 243

\bibitem[Shemmer \& Netzer(2002)]{SN}
Shemmer, O. \& Netzer, H.  2002, \apj, 567, L22

\bibitem[Spitzer(1978)]{Spitzer}
Spitzer, L.  1978, Physical Processes in the Interstellar Medium (New York: Wiley-Interscience), 46

\end{thebibliography}
\end{document}